\documentstyle[11pt,aaspp4]{article}

\def\be{\begin{equation}}
\def\ee{\end{equation}}
\def\ba{\begin{eqnarray}}
\def\ea{\end{eqnarray}}
\def\go{\mathrel{\raise.3ex\hbox{$>$}\mkern-14mu
             \lower0.6ex\hbox{$\sim$}}}
\def\lo{\mathrel{\raise.3ex\hbox{$<$}\mkern-14mu
             \lower0.6ex\hbox{$\sim$}}}

\def\bxi{{\mbox{\boldmath $\xi$}}}
\def\br{{\bf r}}
\def\bv{{\bf v}}

\def\bV{{\bf V}}
\def\cO{{\cal O}}

\begin{document}

\title{GROWTH OF PERTURBATIONS IN GRAVITATIONAL COLLAPSE AND ACCRETION}
\author{Dong Lai}
\affil{Center for Radiophysics and Space Research, 
Department of Astronomy, Cornell University, Ithaca, NY 14853\\
E-mail: dong@spacenet.tn.cornell.edu}
\and
\author{Peter Goldreich}
\affil{California Institute of Technology, 150-21\\
Pasadena, CA 91125\\
E-mail: pmg@gps.caltech.edu}

\begin{abstract}
When a self-gravitating spherical gas cloud collapses or accretes onto 
a central mass, the inner region of the cloud
develops a density profile $\rho\propto r^{-3/2}$ and the velocity approaches
free-fall. We show that in this region, nonspherical perturbations 
grow with decreasing radius. In the linear 
regime, the tangential velocity perturbation increases as $r^{-1}$,
while the Lagrangian density perturbation, $\Delta\rho/\rho$, grows as
$r^{-1/2}$. Faster growth occurs if the central collapsed
object maintains a finite multiple moment, in which case
$\Delta\rho/\rho$ increases as $r^{-l}$, where $l$ specifies the angular 
degree of the perturbation. These scaling relations are different from
those obtained for the collapse of a homogeneous cloud. 
Our numerical calculations indicate that nonspherical perturbations are damped
in the subsonic region, and that they grow and approach the asymptotic 
scalings in the supersonic region. The implications of our results to 
asymmetric supernova collapse and to black hole accretion are briefly
discussed.
\end{abstract}

\keywords{hydrodynamics -- accretion -- supernovae}

\section{Introduction}

Gravitational instability is responsible for a wide range of
structures in the observable universe. Much effort has been devoted to
structure formation in cosmological models, and it is well-established that the
Hubble expansion slows the growth of perturbations (e.g., Peebles 1980).
In this paper we are interested in the behavior of nonspherical perturbations
in the background of spherical collapse or accretion. 

Hunter (1962) studied the collapse of a homogeneous, pressureless,
dust cloud, and showed that in the linear regime, perturbations 
of arbitrary shape and scale grow asymptotically as 
$\delta\rho/\rho\propto (t_0-t)^{-1}\propto R^{-3/2}$, where $t=t_0$ denotes
the time of complete collapse, and $R$ is the cloud radius.
It was thought that this instability might be responsible for the 
fragmentation of collapsing protostellar clouds. Lin, Mestel and Shu (1965)
studied the collapse of a homogeneous ellipsoidal cloud of dust, and showed
that its ellipticity increases until a sheet or pancake forms.
Homogeneity plays a key role in both examples. However, the presence of a
slight initial central concentration significantly alters the evolution of a
cloud. Since the dynamical time, $(G\rho)^{-1/2}$, is shortest in the central
region, a cuspy density and velocity profile develops. As a result, one may
question the applicability of the Hunter and Lin-Mestel-Shu instabilities to 
realistic astrophysical situations. Goodman \& Binney (1983) 
have already commented on problems with the Lin-Mestel-Shu instability for 
inhomogeneous clouds. In this paper, we show that perturbation growth is slowed
by the central mass concentration in an inhomogeneous collapse.

A related problem concerns the stability of spherical accretion flow.
Bondi (1952) found a class of solutions describing steady-state
accretion onto a compact object from a homogeneous medium.
There is a unique transonic ``critical'' flow, for which the mass flux is
maximum; the other subcritical solutions describe
subsonic flows. Bondi speculated that nature would prefer 
the critical flow (see Shu 1992 for a discussion)
and underlined the importance of a linear stability analysis. 
After several attempts by a number of authors, the correct analysis 
was achieved by Garlick (1979) and Moncrief (1980), who showed that  
both the critical and subcritical flows are globally stable. 
However, this does not mean that perturbations can not grow 
spatially. Indeed, we show in this paper that nonspherical perturbations
carried by fluid elements are amplified in the supersonic region 
of the critical flow (see also Goldreich, Lai \& Sahrling 1996;
Kovalenko \& Eremin 1998).  
 
The present study originated from our attempts to understand the origin of
asymmetric supernova (Goldreich, Lai \& Sahrling 1996; hereafter GLS). 
A large body of evidence suggests that Type II supernovae are
globally asymmetric, and that neutron stars receive kick velocities
of order a few hundred to a thousand km~s$^{-1}$ at birth
(see, e.g.,  GLS, Cordes \& Chernoff 1998, and references therein).
The origin of the kick is unknown.
A class of mechanisms relies on local hydrodynamical instabilities in the
collapsed stellar core (e.g., Burrows et al.~1995;
Janka \& M\"uller 1994,~1996; Herant et al.~1994)
that lead to asymmetric matter ejection and/or
asymmetric neutrino emission; but numerical simulations indicate these
instabilities are not adequate to account for kick velocities
$\go 100$~km~s$^{-1}$ (Burrows \& Hayes 1996; Janka 1998). 
Global asymmetric perturbations of presupernova cores may be required
to produce the observed kicks (GLS; Burrows \& Hayes 1996). 
GLS suggested that overstable g-modes driven by shell nuclear
burning might provide seed perturbations which could be amplified during core 
collapse.\footnote{See Lai \& Qian (1998) and Arras \& Lai (1999) for 
discussion/review on alternative mechanisms which rely on asymmetric
neutrino transport induced by strong magnetic fields.}

Hints regarding perturbation growth during collapse are obtained 
by considering the collapse of finite-mass fluid
shells onto central point masses. In the context of core-collapse
supernova, one might imagine that the spherical
shell mimics the outer supersonic region, while the central mass 
represents the homologous core. A large number of growing modes 
can be identified (GLS; Appendix B). In particular, the dominant 
dipole mode grows according to $\Delta R(\theta,t)/R(t)\propto (t_0-t)^s$,
with $s<0$, where $R$ is the mean radius of the shell and
$\Delta R$ is the perturbation. For a core-shell mass of order unity, 
the power-law index $s\simeq -1$, corresponding to $\Delta R/R\propto
R^{-3/2}$. 

The thin-shell model is too simplistic to be applicable to 
real presupernova collapse. One approach is to
determine the stability of the self-similar collapse solution
(Goldreich \& Weber 1980; Yahil 1983). An analysis by Goldreich \& Weber 
(1980) shows that the inner homologous core is stable against nonradial 
perturbations. This is not surprising given the significant role played by 
pressure in the subsonic collapse. Pressure is less important in the
supersonically collapsing region, making it more susceptible to 
large scale instability. 
A stability analysis of Yahil's self-similar solution, 
which extends the Goldreich-Weber solution to include a supersonically 
collapsing outer core, does not reveal any unstable global mode before the 
proto-neutron star forms (Lai 2000).
However, it is perhaps more illuminating to treat the initial-value problem,
and determine how initial
perturbations evolve as the collapse proceeds. We carry out such an analysis 
in this paper. 

The rest of the paper is organized as follows. 
The basic perturbation equations are summarized in \S 2, and in \S 3 we derive
the asymptotic scaling relations for the perturbations in the regime
where the collapse/accretion flow is supersonic. In \S 4 we 
present a numerical study of the evolution of perturbations during collapse;
this numerical study not only confirms the analytic asymptotic relations,
but also explores the regime where pressure is important. 
We discuss the implications of our results in \S 5.

\section{Basic Equations}

We consider barotropic fluid obeying 
the equation of state $p=K\rho^\gamma$, where $\gamma$ is the adiabatic
index and $K$ is a constant. 
The unperturbed flow is spherically symmetric, with velocity
in the radial direction. The Eulerian perturbations
of density, $\rho$, and velocity, $\bv$, can be decomposed into 
different angular modes, each of which has the form
\ba
\delta\rho(\br,t)&=&\delta\rho(r,t)Y_{lm}(\theta,\phi),\\
\delta\bv(\br,t)&=&\delta v_r(r,t)Y_{lm}(\theta,\phi)\,{\hat r}
+\delta v_\perp(r,t)\hat\nabla_\perp Y_{lm}(\theta,\phi)+
\hat\nabla_\perp\times\left[\delta v_{\rm rot}(r,t)
Y_{lm}(\theta,\phi)\hat r\right],
\label{pertv}\ea
where 
\be
{\hat\nabla}_\perp\equiv {\hat\theta}{\partial\over\partial\theta}
+{{\hat\phi}\over\sin\theta}{\partial\over\partial\phi},
\ee
and $\hat r,\,\hat\theta,\,\hat\phi$ are unit vectors in spherical
coordinates. The perturbations of pressure, $p$, and gravitational
potential, $\psi$, have the same angular dependence as $\delta\rho(\br,t)$. 
The perturbed mass continuity equation reads
\be
{d\delta\rho\over dt}+(\nabla\cdot\bv)\delta\rho
+{1\over r^2}\left(r^2\rho\delta v_r\right)'
-l(l+1){\rho\delta v_\perp\over r}=0,\label{dpertrho}
\ee
where $\rho,\,\bv=v\,\hat r$ denote the
unperturbed spherical flow variables, prime stands for $\partial/\partial r$, 
and $d/dt=\partial/\partial t+\bv\cdot\nabla$ is the total
time derivative. The perturbed radial Euler equation can be written as
\be
{d\delta v_r\over dt}+v'\delta v_r=-\left(
{\delta p\over\rho}\right)'-\left(\delta\psi\right)',
\label{dpertvr}
\ee
and the tangential Euler equation reduces to
\ba
&&{d\delta v_\perp\over dt}+{v\over r}\delta v_\perp=-{1\over r}\left(
{\delta p\over\rho}\right)-{1\over r}\delta\psi,\label{dpertvp}\\
&&{d\over dt}(r\delta v_{\rm rot})=0.\label{dpertt}
\ea
The perturbed Poisson equation is 
\be
{1\over r^2}\left[r^2(\delta\psi)'\right]'-{l(l+1)\over
r^2}\delta\psi=4\pi G\delta\rho.
\label{dpertpsi}
\ee

The vorticity perturbation reads
\be
\nabla\times\delta\bv=-{\delta v_T\over r}\,{\hat r}\times
\hat\nabla_\perp Y_{lm}
+{l(l+1)\over r}\delta v_{\rm rot}Y_{lm}\,\hat r
+{1\over r}(r\delta v_{\rm rot})'\,\hat\nabla_\perp Y_{lm},
\ee
where
\be
\delta v_T\equiv \delta v_r-\delta u',
\ee
with 
\be
\delta u(r,t)\equiv r\delta v_\perp(r,t).
\label{durdvperp}
\ee
Thus $\delta v_{\rm rot}$ and $\delta v_T$ are related to the vorticity of the 
perturbed flow.  Equation (\ref{dpertvp}) is transformed to
\be
{d\delta u\over dt}=-{\delta p\over\rho}-\delta\psi,
\label{dpertu}
\ee
with the aid of equation (\ref{durdvperp}). 
Combining equations (\ref{dpertvr}) and (\ref{dpertu}), we find
\be
{d\over dt}\delta v_T=-v' \delta v_T
\label{dpertrot}
\ee

Equations (\ref{dpertt}) and (\ref{dpertrot}) express the conservation of
circulation in a barotropic fluid.\footnote{For homogeneous collapse or 
expansion, as in an expanding universe, we have $v=(\dot R/R)r$, where $R$
is the scale factor. Equation (\ref{dpertrot}) 
then becomes the familiar $d(R\delta v_T)/dt=0$.} 
The former also reflects the conservation of 
angular momentum; following a fluid element, $\delta v_{\rm rot}\propto 1/r$. 
Since, as we shall prove shortly, $v\propto r^{-1/2}$ as $r\to 0$, 
equation (\ref{dpertrot}) implies the asymptotic relation $\delta v_T
\propto r^{1/2}$. We shall focus on irrotational flows from here on. We
neglect $\delta v_{\rm rot}$ because it is 
decoupled from the density perturbation, and $\delta v_T$ because it 
decays inward. The continuity equation (\ref{dpertrho}) for irrotational flow 
simplifies to
\be
{d\delta\rho\over dt}+(\nabla\cdot\bv)\delta\rho
+{1\over r^2}\left(r^2\rho\delta u'\right)'
-l(l+1){\rho\delta u\over r^2}=0.\label{dpertrho2}
\ee

Note that the Poisson equation (\ref{dpertpsi}) has the
following integral solution:
\be
\delta\psi(r,t)=-{4\pi G\over 2l+1}\left[{1\over r^{l+1}}Q_l(r,t)
+r^lS_l(r,t)\right],
\label{pertpsi2}
\ee
where
\be
Q_l(r,t)=\int_0^r\!x^{l+2}\delta\rho(x,t)\,dx,~~~~
S_l(r,t)=\int_r^\infty\!x^{1-l}\delta\rho(x,t)\,dx.
\label{qs}\ee

Equations (\ref{dpertu}), (\ref{dpertrho2}), and (\ref{pertpsi2}) 
determine the perturbed flow. This particular form of the perturbation
equations is convenient for implementation as a Lagrangian 
numerical code (see \S 4).

\section{Asymptotic Analysis}

Consider a cloud in hydrostatic equilibrium with an initial density profile
that decreases outward. As its pressure is depleted, the cloud 
starts to collapse. Since the dynamical time, $(G\rho)^{-1/2}$, decreases 
outward, cuspy density and velocity profiles will be established after 
the core has collapsed. 
By contrast, a uniform density dust cloud collapses
homologously, and remains uniform as the collapse proceeds.
We now study the behavior of the flow and its perturbations in the
asymptotic regime where the gas pressure is negligible compared to gravity. 

\subsection{Unperturbed Spherical Flow}

Consider how the velocity, $v_m$, and density, $\rho_m$, of a fluid
shell with enclosed mass $m$ change as the shell collapses from 
its initial radius $r_{m0}$ to a smaller radius $r_m$. 
The pressure is negligible in the supersonic region, so
for $r_m\ll r_{m0}$, we have 
\be
v_m\simeq -\left({2Gm\over r_m}\right)^{1/2}\propto -r_m^{-1/2}.
\label{vm}\ee
Then from the continuity equation,
we obtain
\be
\rho_m\propto r_m^{-3/2}.
\label{rhom}\ee
Note that these relations do not require the mass of the 
collapsed object to be fixed.  Indeed, with $m_c(t)$ the mass of the 
collapsed core at time $t$, we have 
\be
v(r,t)\simeq -\left[{2Gm_c(t)\over r}\right]^{1/2}\propto r^{-1/2},~~~~
\rho(r,t)\simeq-{\dot m_c(t)\over 4\pi r^2v(r,t)}\propto r^{-3/2},
\label{eulerprofile}
\ee
where $\dot m_c$ is the mass accretion rate onto the 
core.\footnote{If the central region of a pressureless cloud is 
non-singular to begin with, then at the moment when the center reaches infinite
density, the central density and velocity profiles
are $\rho\propto r^{-12/7}$ and $v\propto r^{-1/7}$ (Penston 1969). 
However, after the core has formed, the profiles are given by
equation (\ref{eulerprofile}).}
When the accretion time, $m_c/\dot m_c$, is much longer than the
dynamical time of the flow, equation (\ref{eulerprofile})
describes the inner region of a steady-state Bondi flow.
In a dynamical collapse, when the accreted mass becomes much 
larger than the original core mass, dimensional analysis implies
$m_c(t)=K^{3/2}G^{-(3\gamma-1)/2}t^{4-3\gamma}\bar m_c$,
where $\bar m_c$ is a dimensionless number, and $t$ is measured from
the moment when the center collapses. For $\gamma=1$, 
this reduces to the familiar $m_c=\bar m_c(c_s^3/G)t$ (see Shu 1977). 
Equation (\ref{eulerprofile}) describes
the central region of Shu's expansion-wave solution (Shu 1977), and
the post-collapse extension of the Larson-Penston solution (Larson 1969; 
Penston 1969; Hunter 1977) in the context of star formation, as
well as Yahil's post-collapse solution in the context of core-collapse
supernova (Yahil 1983).

Note that the above asymptotic scaling solution assumes
supersonic flow for $r\rightarrow 0$, i.e., $v \gg
c_s\propto\rho^{(\gamma-1)/2}\propto r^{-3(\gamma-1)/4}$. 
This requires $\gamma<5/3$. The special case of $\gamma=5/3$ is
considered in Appendix A. 

\subsection{Perturbations}

We investigate asymptotic power-law solutions to equations 
(\ref{dpertu}), (\ref{dpertrho2}) and (\ref{pertpsi2}).
Let $\delta\rho\propto r^a$ and $\delta u\propto r^b$. 
The Poisson equation has a general solution of the form
$\delta\psi\sim Q_c/r^{l+1}+r^2\delta\rho$, where the first
term arises from a central multipole moment, 
$Q_c$, and the second term is due to
the density perturbation outside the the central core. In most astrophysical 
situations, $Q_c$ is zero or close to zero; possible exceptions are discussed 
separately in \S 3.3.
For example, in accretion onto a star, the supersonic flow 
may be stopped by a standing shock near the stellar surface. 
Inside the shock, any inhomogeneity carried in by the gas
will be smeared out on a local dynamical timescale.
In accretion onto a Schwarzschild black hole, the event horizon defines
the inner boundary of the supersonic flow, and the no-hair
theorem ensures that mass multipole moments are not
retained by the black hole. Thus we have $\delta\psi
\propto r^{a+2}$. 

In the asymptotic regime, $d/dt\to v(\partial/\partial r)$, so 
equations (\ref{dpertu}) and (\ref{dpertrho2}) reduce to:
\ba
&&{bv\delta u\over r}+\gamma K\rho^{\gamma-2}\delta\rho
+\delta\psi=0,\label{pertdu}\\
&&\left(a+{3\over 2}\right){v\delta\rho\over r}
+\left[b\left(b-{1\over 2}\right)-l(l+1)\right]{\rho\delta u\over r^2}=0.
\label{pertdrho}\ea
Equation (\ref{pertdrho}) implies $b=a+2$ for $a\neq -3/2$. 
Equation (\ref{pertdu}) has the scaling form $\cO(br^{b-3/2})+\cO(K
r^{a-3(\gamma-2)/2})+\cO(r^{a+2})=0$, from which we see immediately that
$b=0$ and $a=-2$ (for $\gamma<5/3$). 
To obtain the scaling
behavior for $\delta v_r=\delta u'$, we need a
higher order correction for $\delta u$. Let $\delta u=\delta u_0
+\delta u_1 r^{b_1}$,
where $\delta u_0$ and $\delta u_1$ are constants independent of
$r$. For $K=0$ (the pressureless case) or
$\gamma<2/3$ so that $\delta p/\rho<\delta\psi$ asymptotically,
equation (\ref{dpertu}) gives $\cO(b_1v\delta u_1 r^{b_1-1})\sim\cO(r^{a+2})$.
We then have $b_1=a+7/2=3/2$. For $K\neq 0$ and $\gamma>2/3$, equation
(\ref{dpertu}) reduces to $\cO(b_1v\delta
u_1r^{b_1-1})+\cO(r^{a-3(\gamma-2)/2})=0$, which gives
$b_1=(5-3\gamma)/2$. To summarize, 
the asymptotic scaling relations for the perturbations are:
\begin{eqnarray}
\delta u &\sim& \delta u_0+\delta u_1 r^{b_1},\label{pertscale1}\\
\delta v_r &\propto& r^{b_1-1},~~~\delta v_\perp\propto
r^{-1},\label{pertscale2}\\
\delta\rho &\sim& -2l(l+1){\rho\delta u\over rv}
\propto r^{-2},\label{pertscale3}
\end{eqnarray}
where 
\be
b_1=\cases{3/2 & for $K=0$ or $\gamma<2/3$,\cr
	(5-3\gamma)/2 & for $K\neq 0$ and $\gamma\ge 2/3$.\cr}
\ee
Note that the above results apply for $\gamma<5/3$.
The special case of $\gamma=5/3$ Bondi accretion is discussed in
Appendix A.

\subsection{Physical Interpretation}

Equations (\ref{pertscale1})-(\ref{pertscale3}) describe the Eulerian
perturbations. To derive scaling relations for the Lagrangian displacement,  
$\bxi(\br,t)=\xi_r(r,t)Y_{lm}\hat r+\xi_\perp(r,t)\hat\nabla_\perp Y_{lm}$, 
and for the Lagrangian density perturbation, $\Delta\rho$, we note that
$\Delta\bv=\delta\bv+(\bxi\cdot\nabla)\bv
={d\bxi/dt}=({\partial\bxi/\partial t})+(\bv\cdot\nabla)\bxi$, which yields
\ba
\delta v_r&=&{\partial\xi_r\over\partial t}+v\xi_r'-v'\xi_r,\\
\delta v_\perp&=&{\partial\xi_\perp\over\partial t}+v\xi_\perp'-{v\over r}
\xi_\perp.
\ea
In the asymptotic regime, we find using equation (\ref{pertscale2}) that
\be
\xi_r\propto r^{b_1+1/2},~~~\xi_\perp\propto r^{1/2}.
\label{scalexi}
\ee
The Lagrangian density and velocity perturbations are
\be
{\Delta\rho\over\rho}\sim {\Delta v_\perp\over v}\propto r^{-1/2},~~~~
{\Delta v_r\over v}\propto r^{b_1-1/2}.\label{pertscale}
\ee
The scaling relations (\ref{scalexi}) and (\ref{pertscale}) can also be
derived directly from Lagrangian perturbation theory. 

Equations (\ref{pertscale1})-(\ref{pertscale3}) and
(\ref{pertscale}) are the main results of this paper.\footnote{This result was 
discussed without derivation in GLS. Similar results were also obtained by 
Kovalenko \& Eremin (1998) in the context of Bondi accretion.} They can be 
understood from the following simple consideration:  
The time that a fluid element spends near radius $r$,
of order $t_r=dt/d\ln r\propto r^{3/2}$,
decreases rapidly as $r$ decreases. Unless the specific torque,
$(\delta\psi+\delta p/\rho)$, increases inward sufficiently rapidly 
to compensate for the decreased time, the specific angular momentum, 
$\delta u$, of the fluid element will be independent of $r$ at small radii.
Thus $\Delta v_{\perp}\propto r^{-1}$. This leads to a tangential
displacement $\xi_\perp\sim t_r\delta v_\perp
\propto r^{1/2}$, which induces a density perturbation $\Delta\rho/\rho\sim 
\xi_\perp/r\propto r^{-1/2}$.

Equation (\ref{pertscale}) indicates that the fractional 
density and tangential velocity perturbations grow in supersonic 
collapse/accretion. The radial velocity perturbation, however, 
can be affected by pressure even in the regime where the pressure has
negligible effect on the unperturbed flow.
We see that $\Delta v_r$ grows with decreasing $r$ 
when $\gamma>1$, and $\Delta v_r/v$ grows only when $\gamma>4/3$
(but recall that the scaling relations apply only for $\gamma<5/3$).

\subsection{Special Cases: Possibility of Faster Growth}

Now consider a hypothetical situation in 
which the inner boundary of the flow is a ``sticky sphere'': Once
a fluid element enters the sphere, it gets stuck on the spot
where it enters. In this case, a finite multipole moment, $Q_c$,
will accumulate at the center (see eq.~[\ref{qs}]).
In the region where $Q_c$ dominates the gravitational perturbation,
we have $\delta\psi\sim Q_c/r^{l+1}\gg r^2\delta\rho$.
It is easy to show from equations (\ref{pertdu})-(\ref{pertdrho})
that the perturbations have the following scaling behavior
(for $\gamma<5/3$):
\begin{eqnarray}
\delta u &\simeq & {2\over 2 l-1}{r\over v}\delta\psi
\propto r^{1/2-l},\label{pertspecial1}\\
\delta\rho &\simeq& -{3\over 2}{\rho\delta u\over r v}\propto
r^{-l-3/2},\label{pertspecial2}\\
\delta\psi &\simeq& -\left({4\pi G\over 2l+1}\right){Q_c\over r^{l+1}}
\propto r^{-l-1}.\label{pertspecial3}
\end{eqnarray}
These scalings depend on $l$. Even for $l=1$, the growth is
faster than the case discussed in \S 3.2 and \S 3.3.
These scalings can be understood as follows: 
The central multipole moment exerts a torque
$\propto r^{-l-1}$ and a radial force $\propto r^{-l-2}$
on a fluid element. The angular momentum grows as
$t_r\delta\psi\propto r^{-l+1/2}$, and the radial velocity perturbation
grows as $\delta v_r\propto r^{-l-1/2}$. Thus we have
$\delta v_r/v\sim\delta v_\perp/v\propto
r^{-l}$. The Lagrangian displacement is $\xi_r\sim\xi_\perp
\sim t_r\delta v\propto r^{-l+1}$, and the resulting density 
perturbation is $\Delta\rho/\rho\sim\xi/r\propto r^{-l}$.

As we have argued in \S 3.2, it is unlikely the scaling relations
derived in this section will apply in general situations
since it is hard to imagine that the central core can retain its
multipole moments for a time much longer than the dynamical time of the
flow at the inner boundary. There are two possible exceptions: 

(i) For $l=1$ modes: The core
has an extra degree of freedom, i.e., it can have linear 
motion in response to flow perturbations. This gives
rise to a net central dipole moment and 
the possibility of faster growth of dipolar perturbations. 
Let $m_c$ be the core mass and $Z_c$ be the position (along the $z$-axis)
of its center-of-mass relative to the the origin of 
coordinates. The displaced core induces a potential perturbation 
$\delta\psi\simeq -(Gm_cZ_c/r^2)\cos\theta$. From equations 
(\ref{pertspecial1})-(\ref{pertspecial3}), the flow perturbations
are given by
\be 
{\delta\rho\over\rho}\simeq {3Z_c\over 2r}\cos\theta,~~~~
{\delta v_r\over v}\simeq {Z_c\over 2r}\cos\theta,~~~~
{\delta v_\perp\over v}\simeq {Z_c\over r}\sin\theta,
\label{pertdipole}\ee
where we have used $v\simeq -(2Gm_c/r)^{1/2}$. 
However, the asymptotic perturbations given by equation (\ref{pertdipole})
simply represent a spherical flow centered at $Z_c$. Thus it is not
surprising 
that the sum of the gravitational force exerted on the core plus the rate
at which momentum flows across a surface surrounding it sum to zero. 

(ii) For $l=2$ modes: Suppose the central core 
consists of a rotating star, or a star with a massive circumstellar disk. 
This could result from an early phase of accretion/collapse with
significant angular momentum. Subsequent spherical accretion, with no
net angular momentum, will be affected by the central quadrupole.

\section{Numerical Calculations of Linear Perturbation Growth}

The asymptotic scaling relations derived in \S 3 apply only
in the supersonic regime. To determine the behavior of 
perturbations under general conditions, 
we numerically follow the collapse of a self-gravitating cloud
and evolve the nonspherical perturbation carried by each fluid element. 
Because of the large disparity in the timescales involved in 
the central region and the outer region, it is essential for the code
to have a wide dynamical range. Previous multidimensional simulations of
Bondi accretion (e.g., Ruffert 1994) did not achieve high enough 
resolution in the central region to reveal 
the growth of perturbation. We restrict our calculations to the linear
regime. Thus perturbations associated with different $Y_{lm}$ evolve
independently, and our calculations involve only one spatial dimension. 

We have constructed a one-dimensional Lagrangian 
finite-difference code. The unperturbed flow variables $(r,v,\rho)$ are 
followed with a standard scheme (Bowers \& Wilson 1991), and
the Eulerian perturbations $(\delta\rho,\delta u,\delta\psi)$
are evolved using equations (\ref{dpertu}), (\ref{dpertrho2}) and
(\ref{pertpsi2}). The flow is covered by a uniform mass grid.
The quantities $r,~v,~\delta u,~\delta\psi$ are zone-edge-centered,
while $\rho,~\delta\rho$ are zone-centered. 
A staggered leapfrog integration scheme is adopted
to ensure second-order accuracy in time.

\subsection{Pressureless Collapse}

To calibrate our code and check the asymptotic scaling relations of \S 3,
we study the collapse of a centrally concentrated dust cloud.
Figure 1 shows an example of such a collapse calculation. 
The cloud, of total mass $m=1$ and radius $r=1$, is initially at rest 
with density profile $\rho\propto r^{-1}$; i.e. the radius of a mass shell
with enclosed mass $m$ is $r_m=m^{1/2}$. We initialize an $l=2$ perturbation 
with $\delta\rho/\rho=1$ in arbitrary units and $\delta u=0$. The Poisson 
equation is solved to give the potential perturbation, $\delta\psi$. 
We impose an inner boundary at $r_c=10^{-3}$: Once a mass shell enters 
this boundary, it is removed from the simulation domain, and its 
perturbations are immediately smeared out; i.e., the central quadrupole 
moment, $Q_c$, is maintained at zero. 
The profiles $\rho\propto r^{-3/2}$ and $v\propto r^{-1/2}$
for the unperturbed flow are established near the center 
as the collapse proceeds. Figure 1 depicts the evolution 
of the perturbation carried by three different mass shells.
We see that as the mass shells collapse to small radii, the analytic 
asymptotic scalings $\delta u\rightarrow$ constant and $\delta\rho/\rho\propto
r^{-1/2}$ are achieved. Calculations with other initial conditions
confirm that these scalings are generic features of perturbation 
growth in the absence of central multipole moments. 

We have also studied the case where the central multipole moment,
$Q_c$, is nonzero, and confirmed the steeper scalings derived in \S 3.4.

\subsection{Collapse with Finite Pressure}

Next we study the collapse of clouds having finite pressure.
For definiteness, we choose the initial cloud to be a $\gamma=4/3$
spherical polytrope in hydrostatic equilibrium. 
The collapse is initiated by reducing $\gamma$ to $1.3$, and by reducing
$K$ by $10\%$. Either one of these reductions alone is adequate to induce the
collapse. This mimics Type II supernova collapse, where
a white dwarf core of a massive star collapses to a neutron star.
However, to focus on the growth of perturbations during the collapse,
we do not include a shock in our calculation. 
We define an inner boundary at $r_c=0.005$; the original cloud 
radius is $r=1$ and its mass $m=1$. When the central density
becomes greater than $10^6$, corresponding to a few times nuclear
density if the initial cloud has mass and radius typical of
a Chandrasekhar mass white dwarf, we cut out the flow inside $r_c$
from the computational domain. This enables us to follow the 
collapse and accretion of the rest of the cloud. 

Figure 2 shows the unperturbed density and velocity of several different mass
shells as functions of their Lagrangian radii. The inner region collapses
homologously; the Mach numbers of individual shells remain below unity outside 
$r_c$. The outer region of the flow goes through a
transonic point, and eventually attains the free-fall asymptotics,
with $v\propto r^{-1/2}$ and $\rho\propto r^{-3/2}$. 
Inspecting the flow profiles at different time slices (not shown), we confirm
that our numerical results agree with the self-similar solution (Yahil 1983) 
in the regime where it applies. 

Figure 3 and Figure 4 show two examples of the evolution of $l=2$ perturbations
during the collapse depicted in Fig.~2. In Fig.~3, the initial
perturbation is chosen to be $\delta\rho/\rho=1$ and $\delta u=0$, while in
Fig.~4, the initial perturbation corresponds to the eigenfunctions of the
first g-mode of a $\gamma=4/3$ polytrope, with adiabatic index $\gamma_1=5/3$.
This value of $\gamma_1$ is used only for setting up
the initial perturbations; after the collapse starts, the adiabatic index 
is set to $\gamma=1.3$. We see that the perturbations carried by the inner mass
shells ($m=0.2,~0.6$), which never become supersonic, vary in an oscillatory
manner with no increase in amplitude. This is consistent with the result 
of Goldreich \& Weber (1980) that the homologous inner core of a collapsing
$\gamma=4/3$ polytrope is stable against non-radial perturbations. 
However, the outer region of the cloud ($m\go 0.85$) attains
a high Mach number and eventually approaches free-fall. 
We see from Figs.~3 and 4 that the density perturbation grows in this outer 
region, and that the asymptotic scaling relations derived in \S 3.2 are 
recovered.

\section{Discussion}

Nonspherical perturbations are 
amplified during the supersonic collapse or accretion of a centrally 
concentrated gas cloud. We have derived asymptotic scaling 
relations for their growth.
These general results have implications for several different
astrophysical problems which we now discuss.

In spherical accretion onto a black hole, we expect 
the radiative efficiency to increase as a result of nonradial 
perturbations in the flow. In the asymptotic regime, 
the tangential velocity perturbation scales as $\delta v_\perp\propto
r^{-1}$, and the corresponding Mach number scales as $\delta v_\perp/c_s
\propto r^{(3\gamma-7)/4}$. For $\gamma<5/3$, the Mach number grows 
faster than $r^{-1/2}$. Similarly, the Mach number associated with the
radial velocity perturbation (see eq.~[\ref{pertscale2}] for $\gamma>2/3$)
scales as $\delta v_r/c_s\propto r^{-3(\gamma-1)/4}$. One might expect the 
formation of shocks which 
lead to thermalization of the flow and high radiative efficiency (see 
Chang \& Ostriker 1985 for previous discussion on the formation of shocks in
spherical flows). This point has also been noted recently by Kovalenko \&
Eremin (1998), who derived similar scaling relations for Bondi accretion. 

In the context of core-collapse supernovae, our results indicate
that perturbations in the homologous inner core do not grow, but
those in the outer core, involving $\sim 15\%$ of the core mass, 
and in the envelope are amplified. Since $\delta\rho/\rho$ scales as
$r^{-1/2}$, we expect 
that the amplification factor is at most $10$ for $r$ decreasing from 
$1500$~km to $15$~km. It is possible that dipole perturbations obey the
$r^{-1}$ scaling (see \S 3.4), in which case the amplification factor
could be larger. Interestingly, if overstable g-modes driven by shell
nuclear burning are responsible for the seed of presupernova perturbations 
(see GLS), it is exactly at the outer core where the perturbation amplitude
is expected to be the largest. The asymmetric density perturbation
may lead to asymmetric shock propagation and breakout, which then give rise
to asymmetry in the explosion and a kick to the neutron star 
(e.g., Burrows \& Hayes 1996).

Finally, we note that our analysis has neglected rotation (i.e., the flow 
does not have a net angular momentum). In the perturbative regime, 
rotation (around the $z$-axis) is represented by the last term of
equation (\ref{pertv}) with $m=0$, i.e., $\delta {\bf v}_{\rm rot}
=-\delta v_{\rm rot}(r,t)(\partial Y_{l0}/\partial\theta)\,\hat\phi$.
As shown in \S 2, this rotational perturbation is decoupled from 
the density perturbation. Therefore, provided the rotational velocity 
is small in comparison to the radial velocity,
i.e., $|\delta v_T|\ll |v|$, we expect our scaling relations for the growth of
perturbations to be valid.

\acknowledgments

A major portion of this research was done between 1995 and 1997, when D.L. 
was a postdoc in theoretical astrophysics at Caltech; support from a
Richard C. Tolman fellowship is gratefully acknowledged. 
This work is supported in part by NASA grant NAG 5-8356,
by a research fellowship from the Alfred P. Sloan foundation, 
and by NSF grant 94-14232.

\appendix

\section{Perturbation of Bondi Accretion for $\gamma=5/3$}

The Bondi solution for $\gamma=5/3$ is special because the flow remains
subsonic for $r>0$. We can think of the sonic point as being located at $r=0$
(in Newtonian theory). For $r\ll GM/c_\infty^2$, where $M$ is the central mass
and $c_\infty$ is the sound speed at infinity, the flow velocity and
density are given by
\ba
|v|&\simeq& c_s\simeq 
\left({GM\over 2r}\right)^{1/2},\\
\rho &\simeq &\rho_\infty\left({GM\over 2c_\infty^2}\right)^{3/2}r^{-3/2}. 
\ea
Although these have the same scalings as equations (\ref{vm})
and (\ref{rhom}), the asymptotic behavior of the perturbations is
quite different. Indeed, for $\gamma=5/3$, eqs.~(\ref{pertdu}) and
(\ref{pertdrho}) yield
\be
\delta\rho\propto r^{l-1/2},~~~
\delta u\propto r^{l+3/2},~~~
\delta\psi\propto r^l.
\ee
Thus all perturbations decrease as $r$ decreases in keeping with the subsonic
nature of the unperturbed flow. 
If a central multipole 
moment is present (see \S 3.4), the following asymptotics become
dominant:
\be
\delta\rho\propto r^{-l-3/2},~~~
\delta u\propto r^{-l+1/2},~~~
\delta\psi\propto r^{-l-1}.
\ee


\section{Perturbations in Collapsing Spherical Shells}

Consider a spherical fluid shell of mass $M_s$ falling from infinity onto 
a central point mass $M_c$. The shell radius, $R_0(t)$, evolves in time 
according to 
\be 
{d^2R_0\over dt^2}=-{GM\over R_0^2}, ~~~~{\rm with}~~
M=M_c+{M_s\over 2},
\ee
which gives
\be
R_0(t)=\left({9GM\over 2}\right)^{1/3}(-t)^{2/3},
\label{R0}\ee
where we have set $t=0$ at the point of complete collapse. 
The surface density and radial velocity are given by
\be
\Sigma_0(t)={M_s\over 4\pi R_0^2}\propto (-t)^{-4/3},
~~~~~~
V_0(t)={dR_0\over dt}\propto (-t)^{-1/3}.
\label{Sigma0}\ee

\subsection{Perturbation Equations}

The dynamical variables for a perturbed shell
are its surface density $\Sigma(\theta,t)$, radius $R(\theta,t)$,
radial velocity $V_r(\theta,t)=\dot R$, where dot indicates 
$\partial/\partial t$, and tangential velocity $\bV_\perp(\theta,t)$.
We assume spherical coordinates and axisymmetry so that there is no 
$\phi$-dependence.
The core mass, $M_c$, is free to move. We use 
$Z_c(t)$ to denote its displacement from the coordinate origin.
Note that $\Sigma,~V_r,~\bV_\perp$
are rigorously defined from the three-dimensional fluid variables via
\be
\Sigma\equiv {1\over R^2}\int_{R-}^{R+}\!\!\rho\,r^2 dr,~~~~~
V_r\equiv {1\over \Sigma R^2}\int_{R-}^{R+}\!\!\rho\,r^2v_r dr,~~~~~
\bV_\perp\equiv {1\over \Sigma R^2}\int_{R-}^{R+}\!\!\rho\,r^2\bv_\perp dr,
\ee
where the integration runs through the thickness of the shell. 
Using these definitions and the standard hydrodynamical equations,
we derive the continuity and Euler equations for the shell:
\ba
&&{\partial\Sigma\over\partial t}=-{2\dot R\over R}
\Sigma-\Sigma\,\nabla_\perp\cdot\bV_\perp,\label{cont}\\
&&{\partial V_r\over\partial t}=\ddot R=
-{1\over 2}\left[\left({\partial\Phi\over
\partial r}\right)_{R+}+\left({\partial\Phi\over \partial r}\right)_{R-}
\right],\label{radial}\\
&&{\partial\bV_\perp\over\partial t}=-{\dot R\bV_\perp\over R}
-{1\over 2}
\left[\left(\nabla_\perp\Phi\right)_{R+}+
\left(\nabla_\perp\Phi\right)_{R-}\right],
\label{trans}
\ea
where we have assumed that the nonspherical perturbation is small. Note that
here $\nabla_\perp\equiv (1/R)\hat\nabla_\perp=
(1/R)\left[\hat\theta(\partial/\partial\theta)
+(\hat\phi/\sin\theta)(\partial/\partial\phi)\right]
=(1/R)\hat\theta(\partial/\partial\theta)$. The gravitational
potential $\Phi=\Phi_c+\Phi_s$ includes contributions from the core,
$\Phi_c$, as given by
\be
\Phi_c(\br,t)=-{GM_c\over |\br-Z_c\hat z|}=-GM_c\sum_l{Z_c^l\over r^{l+1}}
\left({4\pi\over 2l+1}\right)^{1/2}Y_{l0}(\theta,\phi), 
\ee
for $R>Z_c$.
The potential produced by the shell satisfies
\be
\nabla^2\Phi_s(\br,t)=4\pi G\Sigma(\theta,t)\delta[r-R(\theta,t)].
\label{poisson}
\ee
Finally we need the equation of motion for the core mass:
\be
\ddot Z_c=-\left({\partial\Phi_s\over\partial r}\right)_{r=Z_c,\theta=0}.
\label{zdot}\ee

Consider linear perturbation modes associated with spherical
harmonics $Y_{l0}$:
\ba
&&R(\theta,t)=R_0(t)\left[1+a_l(t)Y_{l0}\right],\\
&&\bV_\perp(\theta,t)=\dot R_0(t)\,b_l(t)\hat\nabla_\perp Y_{l0},\\
&&\Sigma(\theta,t)=\Sigma_0(t)\left[1+c_l(t)Y_{l0}\right].
\ea
Solving equation (\ref{poisson}), the shell potential to linear order in 
$a_l,~c_l$ is 
\be
\Phi_s(\br,t)=-{GM_s\over r}-{1\over 2l+1}\left({GM_s\over R_0}
\right)\left({R_0\over r}\right)^{l+1}\Bigl[c_l+(l+2)a_l\Bigr]Y_{l0},
\ee
for $r>R(\theta,t)$ (outside the shell), and
\be
\Phi_s(\br,t)=-{GM_s\over R_0}-{1\over 2l+1}\left({GM_s\over R_0}
\right)\left({r\over R_0}\right)^l\Bigl[c_l-(l-1)a_l\Bigr]Y_{l0},
\ee
for $r<R(\theta,t)$ (inside the shell). Using the unperturbed
solution (eqs.~[\ref{R0}] and [\ref{Sigma0}]), the perturbation equations
(\ref{cont})-(\ref{trans}) for the shell reduce to
\ba
&& 2t\dot a_l+t\dot c_l-{2\over 3}l(l+1)b_l=0,\label{shell1}\\
&&9t^2\ddot a_l+12 t\dot a_l-2 a_l=-{4M_c\over M}
\delta_{l1}z_c+{4M_c\over M}a_l-{2M_s\over (2l+1)M}\Bigl[
l(l-1)a_l+{1\over 2}c_l\Bigr],\label{shell2}\\
&&3t\dot b_l+b_l={M_c\over M}\delta_{l1}
z_c+{M_s\over (2l+1)M}\Bigl(c_l+{3\over 2}a_l\Bigr),\label{shell3}
\ea
where $\delta_{l1}$ is Kronecker delta, and
\be
z_c\equiv \left({4\pi\over 3}\right)^{1/2}\!
\left({Z_c\over R_0}\right).
\ee
The equation of motion for the core mass, equation (\ref{zdot}), becomes
\be
9t^2\ddot z_c+12t\dot z_c-2z_c={2M_s\over 3M}c_1.
\label{shell4}\ee
Setting 
\be
a_l,~b_l,~c_l,~z_c\propto (-t)^s,
\ee
equations (\ref{shell1})-(\ref{shell4}) reduce to a set of algebraic 
equations from which the eigenvalue, $s$, and the corresponding eigenmode
can be determined. We discuss these eigenmodes below.
 
\subsection{$l=1$ Modes}

There are six roots for $s$. Two of these, $s=1/3,\,-2/3$, are trivial
modes which do not involve any surface density perturbation ($c_1=0$), and
for which the core experiences no acceleration ($\ddot Z_c=0$). These
correspond to the collapse of a uniform shell onto a displaced core; for
$\Delta R =R-R_0\propto (-t)$, $\Delta R/R_0\propto (-t)^{1/3}$, and for 
$\Delta R=$ constant, $\Delta R/R_0\propto (-t)^{-2/3}$.

Two of the remaining four roots correspond to stable modes ($s>0$). 
The two unstable modes are shown in Fig.~5. Both lead to the growth 
of the separation between the center of mass of the shell and the position
of the core: (i) {\it ``Bending'' Mode}: the surface density
perturbation grows because one side of the shell collapses faster than the 
other, and the geometric center of the shell moves in opposition to the
motion of the central mass; when $M_s\rightarrow 0$, the mode has $s=-1$ and
$(a_1,b_1,c_1)=(1,0,-2)$;
(ii) {\it ``Jeans'' Mode}: the surface density perturbation grows due to the 
internal tangential flow in the shell,
while the shell's geometric center suffers little displacement
with respect to the position of the central mass; 
when $M_s\rightarrow 0$, the mode has $s=-1/3$ and
$(a_1,b_1,c_1)=(0,1,-4)$. The ``bending'' mode grows more rapidly, with $s$ 
ranging from $-1$ for $M_s\rightarrow 0$ to
$s=-(\sqrt{17}+1)/6=-0.854$ for $M_c\rightarrow 0$.  

\subsection{$l=2$ Modes}

For $l=2$, the core mass experiences no acceleration. There are two types
of unstable modes as shown in Fig.~6: (i) {\it ``Bending'' Mode}: 
the north pole and south pole of the shell collapse faster and have higher
density than the equator, leading to a quadrupolar density perturbation; 
(ii) {\it ``Jeans'' Mode}: the density perturbation is mainly due to 
tangential fluid motion within the shell. The ``bending'' mode is the more 
rapidly growing mode, with $s$ in the range between $-1$ and $-0.85$;

\subsection{Large-$l$ Limit}

Our results for general $l$ will not be presented here. But it is of 
interest to consider the $l\gg 1$ limit.

(i) {\it ``Bending'' Mode}: The dispersion relation of bending waves
on a pressureless, nonrotating surface is
$\omega^2=2\pi G\Sigma_0|k|$, with $|k|\simeq l/R_0$. This gives
\be
\omega=\pm {s_l\over t},~~~~{\rm with}~~
s_l\equiv {1\over 3}\left({lM_s\over M}\right)^{1/2}.
\ee
The wave evolves as $\exp(i\int\!\omega\,dt)=(-t)^{\pm is_l}$. To obtain
the amplitude evolution, we use the conservation of wave action (energy
per unit mass divided by frequency) $\propto \omega|\Delta R|^2$, which
gives $|\Delta R|\propto (-t)^{1/2}$, and
$|\Delta R|/R_0\propto (-t)^{-1/6}$. Thus
\be
{\Delta R\over R_0}\propto (-t)^s,~~~~~s=-{1\over 6}\pm is_l.
\label{bending}\ee
This agrees with our numerical results.

(ii) {\it ``Jeans'' Mode}: 
The dispersion relation of density wave is $\omega^2=-2\pi G\Sigma_0|k|$, which
gives $\omega=\pm is_l/t$. Similar to (i), we find
\be
{\Delta R\over R_0}\propto (-t)^{-1/6}\exp(i\!\int\!\omega\,dt)
\sim (-t)^s,~~~~~s=-{1\over 6}\pm s_l.
\ee
This also agrees with our numerical results.

\subsection{Effect of Internal Pressure}

We can include the effect of fluid pressure by adding 
a term $-(1/\Sigma)\nabla_\perp(\Sigma C_s^2)$ to the tangential Euler
equation (\ref{trans}); the radial equation (\ref{radial}) is not affected
by pressure. For definiteness, we parametrize
the shell-averaged sound speed, $C_s$, by
\be
C_s=\beta\left({GM\over R_0}\right)^{1/2},
\ee
where $\beta$ is a constant. This amounts to adding a term $-\beta^2c_l$
to the right-hand-side of equation (\ref{shell3}). 

Figures 5 and 6 show the effect of pressure on the eigenvalues of the $l=1$ and
$l=2$ modes, respectively. We see that pressure always tends to stablize the
modes. However, for $M_s/M_c$ less than a few, the ``bending'' mode is
only slightly affected. 

In the large-$l$ limit, the ``bending'' mode is unaffected by the pressure,
thus equation (\ref{bending}) still applies. For the ``Jeans'' mode, the 
dispersion relation is $\omega^2=k^2C_s^2-2\pi G\Sigma_0|k|$, which gives
\be
\omega^2=\left(2\beta^2l^2-{lM_s\over M}\right){1\over 9t^2}.
\ee
For $l\gg M_s/(2\beta^2M)$, we have
$\omega\simeq \pm (\sqrt{2}/3)\beta l/t$. Using similar precedure as
in \S B.4, we obtain
\be
{\Delta R\over R_0}\propto (-t)^s,~~~~~{\rm with}~~s\simeq -{1\over 6}\pm
i{\sqrt{2}\over 3}\beta l.
\ee


\bigskip
\bigskip
\bigskip

\begin{figure}
\plotone{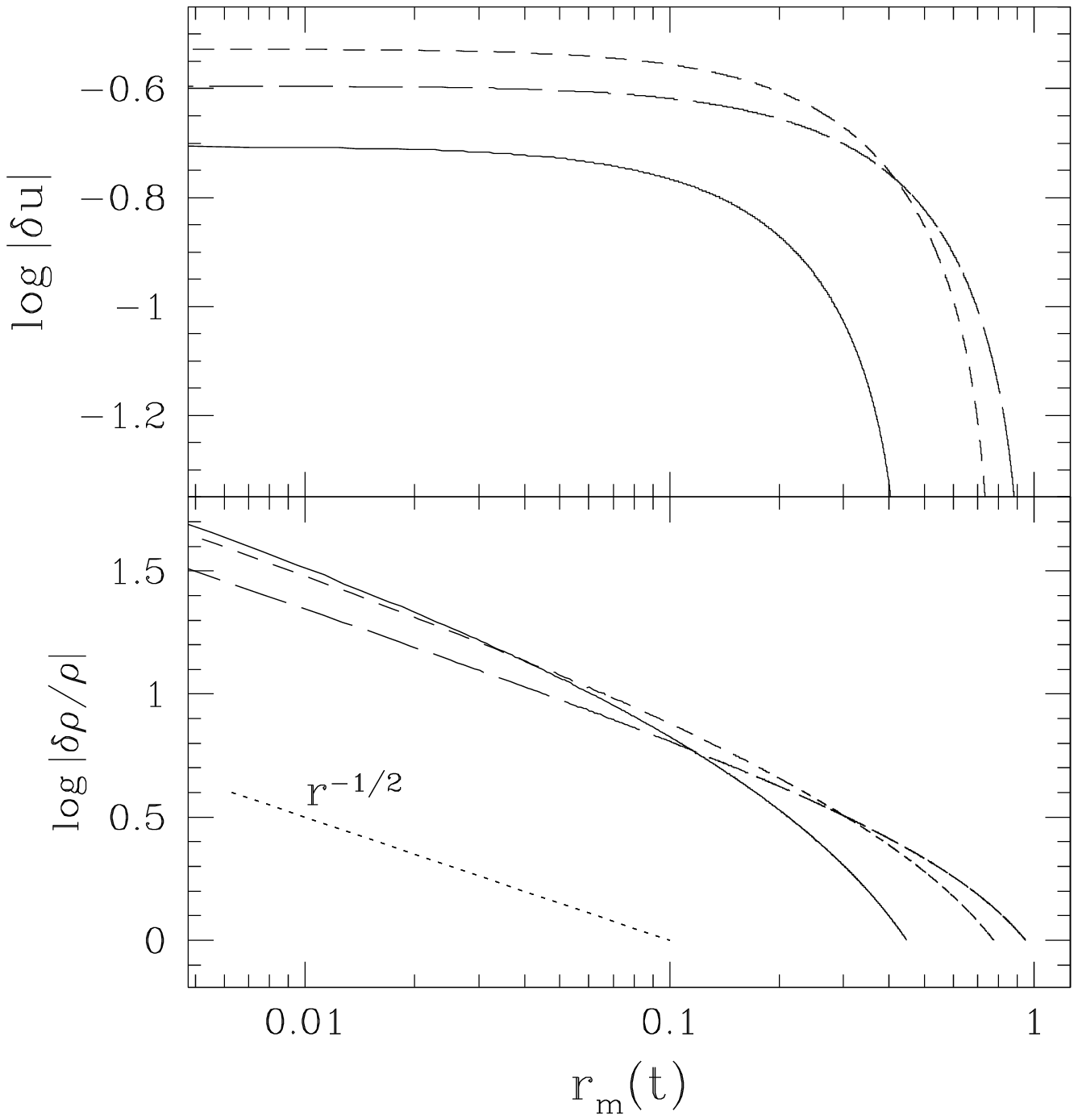}
\caption{
Evolution of an $l=2$ perturbation during the collapse of 
a centrally concentrated dust sphere. 
The velocity potential perturbation, $\delta u$, in the upper panel, 
and the fractional density perturbation, $\delta\rho/\rho$, in the lower 
panel, are plotted against the Lagrangian radius, $r_m(t)$,
for three different mass shells: $m=0.2$ (solid lines),
$m=0.6$ (short-dashed lines), and $m=0.9$ (long-dashed lines).
The dotted line in the lower panel shows the asymptotic scaling relation as 
derived in \S 3.2.
\label{fig1}}
\end{figure}

\begin{figure}
\plotone{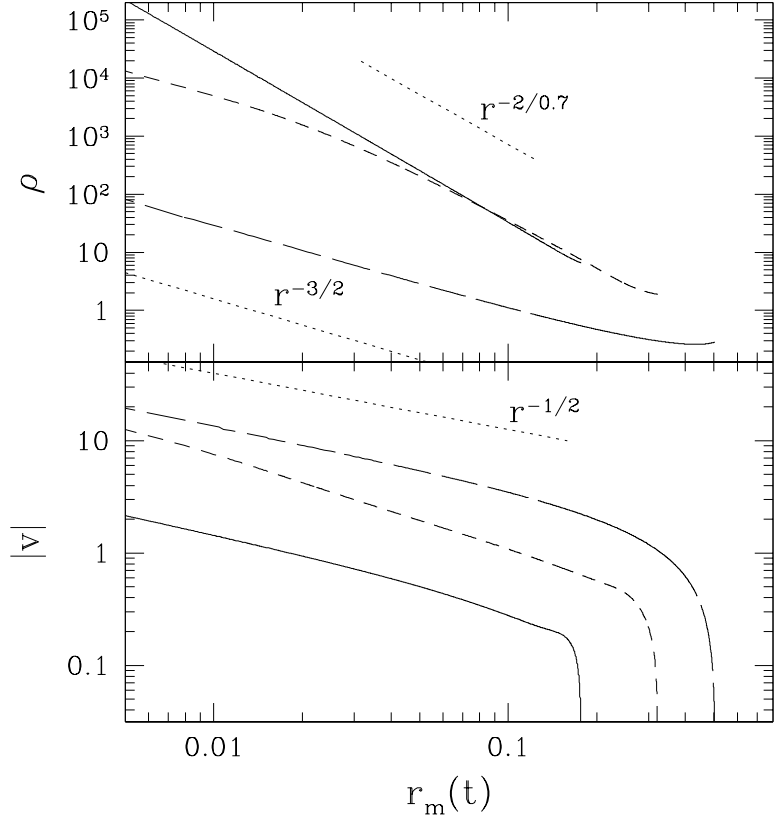}
\caption{
Evolution of the unperturbed flow during the collapse of 
a pressure-depleted $\gamma=4/3$ polytrope. The adiabatic index 
is reduced to $\gamma=1.3$ after the collapse starts.
The density and velocity are plotted against 
the Lagrangian radius, $r_m(t)$, for
of three different mass shells: $m=0.2$ (solid lines),
$m=0.6$ (short-dashed lines), and $m=0.9$ (long-dashed lines).
The dotted lines show the asymptotic scaling relations. Note that
$\rho\propto r^{-2/(2-\gamma)}$ applies to the outer region of
the self-similar flow (Yahil 1983), while $\rho\propto r^{-3/2}$ and
$v\propto r^{-1/2}$ apply to the inner region of the post-collapse flow.
\label{fig2}}
\end{figure}

\begin{figure}
\plotone{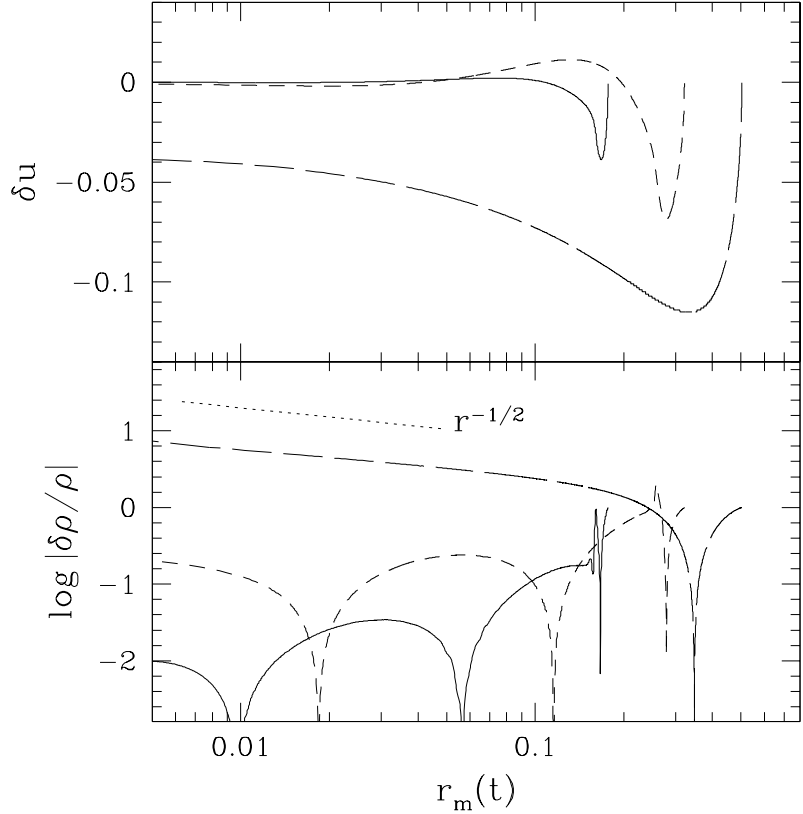}
\caption{
Evolution of the $l=2$ perturbation during the collapse of 
a pressure-depleted $\gamma=4/3$ polytrope. The unperturbed
flow is depicted in Fig.~2. The velocity potential perturbation, $\delta u$,
in the upper panel, and the fractional density perturbation, $\delta\rho/\rho$,
in the lower panel, are plotted against the Lagrangian radius, $r_m(t)$,
for three different mass shells: $m=0.2$ (solid lines),
$m=0.6$ (short-dashed lines) and $m=0.9$ (long-dashed lines).
The initial perturbation is chosen to be
$\delta\rho/\rho=1$ and $\delta u=0$. 
The dotted line shows the asymptotic scaling derived in \S 3.2. 
\label{fig3}}
\end{figure}

\begin{figure}
\plotone{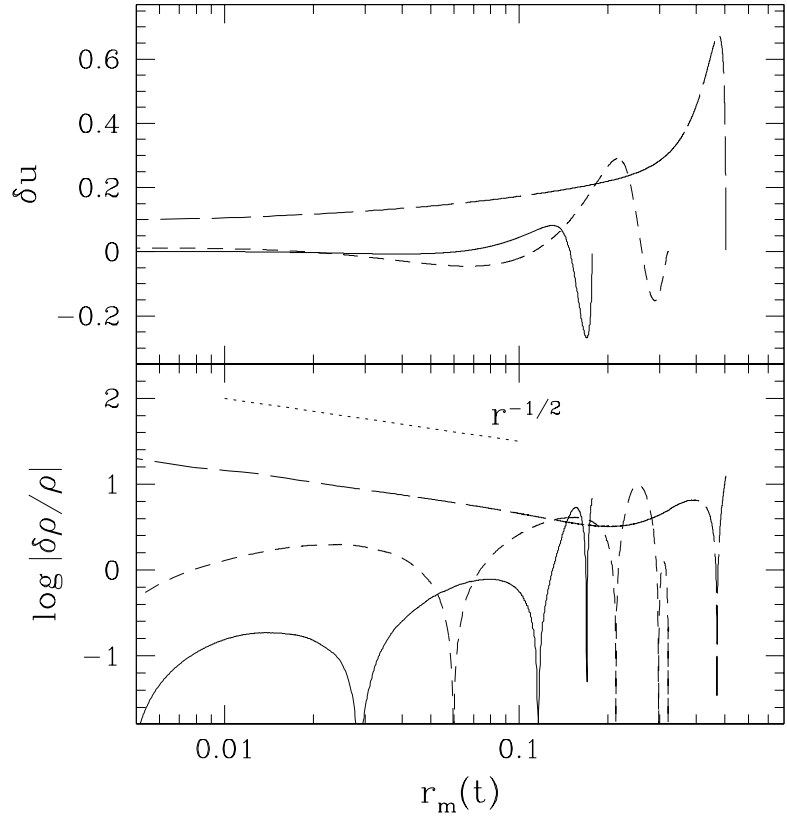}
\caption{
Same as Fig.~3, except that the initial perturbation 
is chosen to correspond to the $g_1$ mode of a $\gamma=4/3$ polytrope with 
adiabatic index $\gamma_1=5/3$.
\label{fig4}}
\end{figure}

\begin{figure}
\plotone{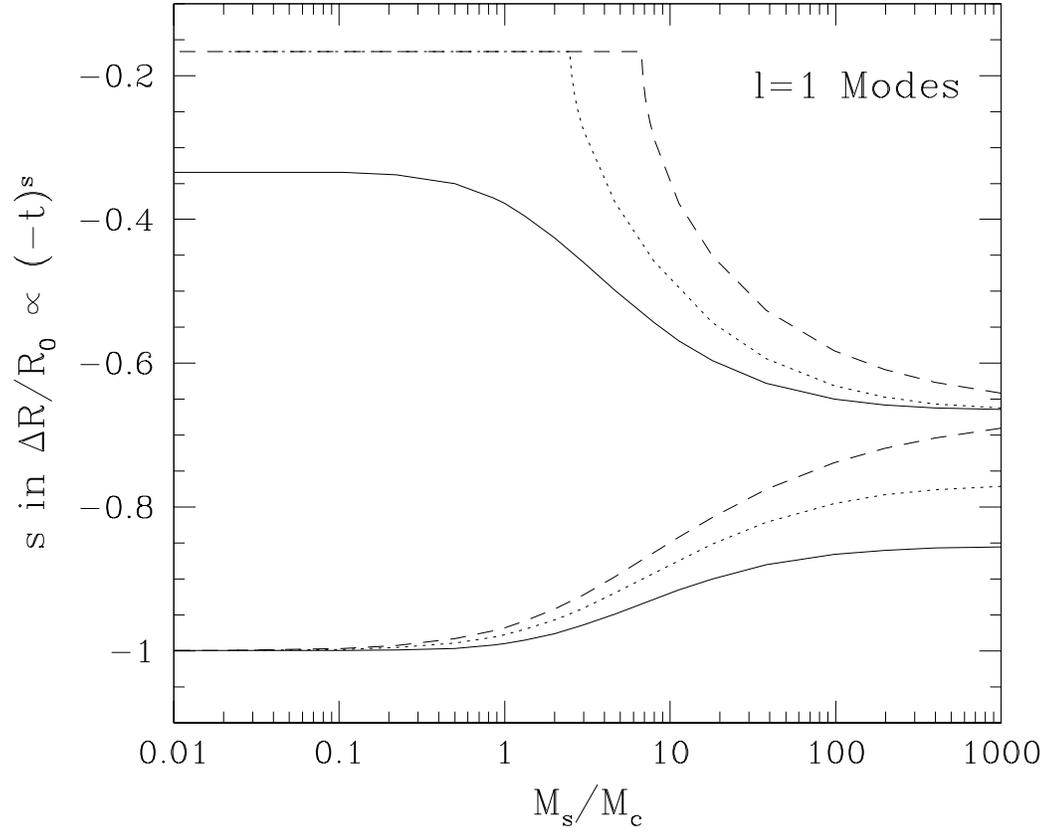}
\caption{
Perturbation modes ($l=1$) in a collapsing shell. The solid 
curves are for zero pressure ($\beta=0$); the lower curve corresponds to the 
``bending'' mode, and the upper to the ``Jeans'' mode. The dotted and dashed 
curves include pressure with $\beta=0.5$ and $\beta=\sqrt{0.5}$, respectively.
Note that when $s$ is complex, only its real part is plotted.
\label{fig5}}
\end{figure}

\begin{figure}
\plotone{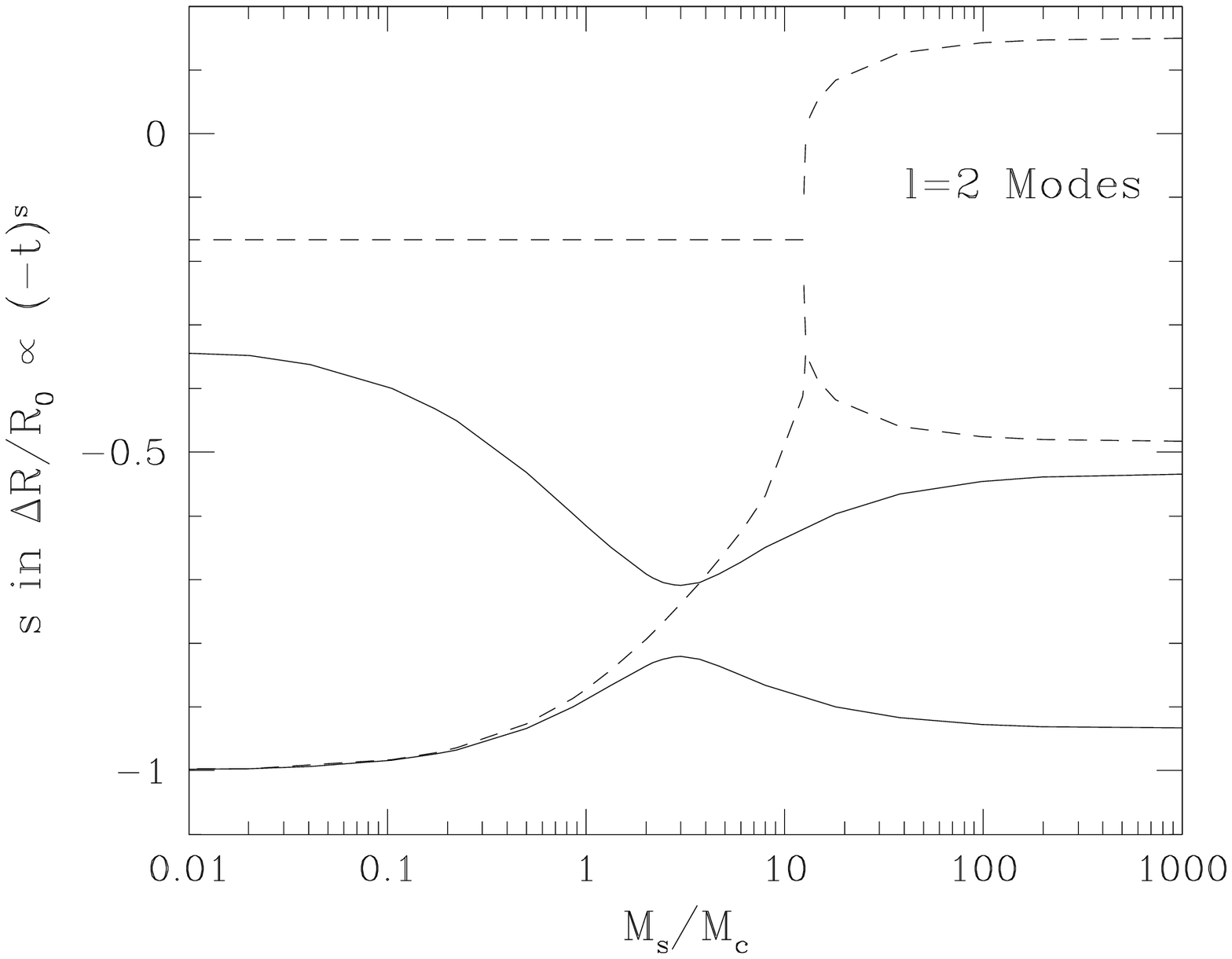}
\caption{
Perturbation modes ($l=2$) in a collapsing shell. The solid 
curves are for zero pressure ($\beta=0$); the lower curve corresponds to the 
``bending'' mode, and the upper corresponds to the ``Jeans'' mode. The dashed 
curves include pressure with $\beta=\sqrt{0.5}$. Note that except for the
lower-left branch, the dashed curves depict only the real part of $s$. 
\label{fig6}}
\end{figure}


\begin{thebibliography}{}


\bibitem[]{}
Arras, P., \& Lai, D. 1999, Phys. Rev. D., in press (astro-ph/9811371).



\bibitem[Bondi 1952]{Bondi52}
Bondi, H. 1952, MNRAS, 112, 195.

\bibitem[]{}
Bowers, R.~L., \& Wilson, J.~R. 1991, Numerical Modeling in Applied
Physics and Astrophysics (Jones \& Bartlett Publishers: Boston).

\bibitem[]{}
Burrows, A., Hayes, J., \& Fryxell, B.~A. 1995, ApJ, 450, 830.

\bibitem[]{}
Burrows, A., \& Hayes, J. 1996, Phys. Rev. Lett., 76, 352.

\bibitem[]{}
Chang, K.~M., \& Ostriker, J.~P. 1985, ApJ, 288, 428.

\bibitem[]{}
Cordes, J.~M., \& Chernoff, D.~F. 1998, ApJ, 505, 315.

\bibitem[]{}
Garlick, A.~R. 1979, A\&A, 73, 171.

\bibitem[]{}
Goldreich, P., Lai, D., \& Sahrling, M. 1996, in ``Unsolved
Problems in Astrophysics'', ed. J.~N. Bahcall and J.~P. Ostriker
(Princeton University Press) (GLS).

\bibitem[]{}
Goldreich, P., \& Weber, S.~V. 1980, ApJ, 238, 990.

\bibitem[]{}
Goodman, J., \& Binney, J. 1983, MNRAS, 203, 265.

\bibitem[]{}
Herant, M., Benz, W., Hix, J., Colgate, S.~A., 
\& Fryer, C. 1994, ApJ, 395, 642.

\bibitem[]{}
Hunter, C. 1962, ApJ, 136, 594.

\bibitem[]{}
Hunter, C. 1977, ApJ, 218, 834.


\bibitem[]{}
Janka, H.-T. 1998, in Proc. of the 4th Ringberg Workshop on Neutrino
Astrophysics (Munich, MPI).

\bibitem[]{}
Janka, H.-Th., \& M\"uller, E. 1994, A\&A, 290, 496.

\bibitem[]{}
Janka, H.-T., \& M\"uller, E. 1996, A\&A, 306, 167.


\bibitem[]{}
Kovalenko, I.~G., \& Eremin, M.~A. 1998, MNRAS, 298, 861.

\bibitem[]{}
Lai, D. 2000, ApJ, submitted.

\bibitem[]{}
Lai, D., \& Qian, Y.-Z. 1998, ApJ, 505, 844.

\bibitem[]{}
Larson, R.~B. 1969, MNRAS, 145, 271.

\bibitem[]{}
Lin, C.~C., Mestel, L., \& Shu, F.~H. 1965, ApJ, 142, 1431.

\bibitem[]{}
Moncrief, V. 1980, ApJ, 235, 1038.


\bibitem[]{}
Peebles, P.~J.~E. 1980, The Large-Scale Structure of The Universe
(Princeton University Press).

\bibitem[]{}
Penston, M.~V. 1969, MNRAS, 144, 425.


\bibitem[]{}
Ruffert, M. 1994, ApJ, 427, 342.

\bibitem[]{}
Shu, F.~H. 1977, ApJ, 214, 488.

\bibitem[]{}
Shu, F.~H. 1992, The Physics of Astrophysics: Gas Dynamics
(University Science Books, Mill Valley, CA), p.80.



\bibitem[]{}
Yahil, A. 1983, ApJ, 265, 1047.

\end{thebibliography}
\end{document}